\documentclass[aps,prb,twocolumn]{revtex4}
\newcommand{\garnet}{\textnormal{{Mn}$_{3}$Al$_{2}$Ge$_{3}$O$_{12}$~}}
\newcommand{\vect}[1]{\ensuremath{\mathbf{#1}}}

\usepackage{graphicx}
\usepackage{epstopdf}

\begin{document}

\title{Experimental study of antiferromagnetic resonance in noncollinear antiferromagnet \garnet{}}

\author{Yu. V. Krasnikova}
\email{krasnikova.mipt@gmail.com}
\affiliation{P. L. Kapitza Institute for Physical Problems RAS, 119334, Moscow, Russia}
\affiliation{National Research University 'Higher School of Economics', Moscow, Russia, 101000, Moscow, Russia}

\author{V. N. Glazkov}
\affiliation{P. L. Kapitza Institute for Physical Problems RAS, 119334, Moscow, Russia}
\affiliation{National Research University 'Higher School of Economics', Moscow, Russia, 101000, Moscow, Russia}

\author{T. A. Soldatov}
\affiliation{P. L. Kapitza Institute for Physical Problems RAS, 119334, Moscow, Russia}
\affiliation{Moscow Institute of Physics and Technology, 141700, Dolgoprudny, Russia}

\begin{abstract}
We have measured antiferromagnetic resonance (AFMR) frequency-field dependences for aluminum-manganese garnet \garnet{} at frequencies from 1 to 125 GHz and at the  fields up to 60 kOe. Three AFMR modes were observed for all orientations, their zero field gaps are about 40 and 70 GHz. Andreev-Marchenko hydrodynamic theory \cite{Marchenko} well describes experimental frequency-field dependences. We have observed hysteresis of resonance absorption as well as history dependence of resonance absorption near gap frequencies below 10 kOe in all three measured field orientations, which are supposedly due to the sample domain structure. Observation of the AFMR signal at the frequencies from 1 to 5 GHz allows to estimate repulsion of nuclear and electron modes of spin precession in the vicinity of spin-reorientation transition at $\vect{H}\parallel[100]$.
\end{abstract}

\maketitle

\section{Introduction}
The compound \garnet orders at $T_N=6.5$K antiferromagnetically \cite{Prandl} with complex twelve-sublattices noncollinear order below the Neel temperature. Crystal symmetry group of this compound is $O_h^{10}$, in ordered phase spins are confined to one of the (111)-type planes. \cite{Plahtii} As there are four equivalent directions of [111] type in the cubic crystal, four antiferromagnetic domains could exist below the Neel temperature. Equivalence of these domains would be broken by an applied magnetic field. The existence of such domains lead to magnetosriction peculiarities of  \garnet{} crystals. \cite{Kazey} Oscillations of these domains boundaries were proposed  \cite{Tikhonov} to explain the nonlinear absorption of RF-signal in Mn-NMR experiments.

Dynamics of magnetic excitations in \garnet{} was studied by inelastic neutron scattering \cite{Plahtii}, antiferromagnetic resonance \cite{Prozorova} and ${}^{55}$Mn NMR.\cite{TikhonovJL} Spin waves dispersion curves were determined in Ref.\onlinecite{Plahtii}. Besides of acoustic modes spin waves spectra include optical (exchange) modes with the gap about 250 GHz. Low-frequency dynamic was studied in details by AFMR at the frequencies from 20 to 80 GHz and at the fields up to 20 kOe.\cite{Prozorova} ${}^{55}$Mn NMR for \garnet{} was studied at the frequencies from 200 to 640 MHz.\cite{TikhonovJL}

Frequency-field dependences of AFMR and NMR in \garnet{} were successfully described by Andreev-Marchenko theory. \cite{Marchenko, Prozorova, Udalov, TikhonovJL} This theory predicts softening (zeroing of the frequency) of one of the AFMR modes at the field of spin-reorientation transition at $\vect{H}\parallel[100]$. In this case AFMR mode should cross NMR mode and hyperfine interaction on manganese ions would lead to the repulsion of these modes. Existence of this interaction was observed in NMR experiments of Ref.\onlinecite{TikhonovJL} as softening of certain NMR modes. Dynamic of the nuclear subsystem and its interaction with the electron subsystem for \garnet{}  was described in details in Ref.\onlinecite{Udalov}.

The aim of our present work was to study spin dynamics in \garnet{} in wide range of frequencies and fields, to study in details frequency-field dependence of AFMR near spin-reorientation transition and also to study domain structure of the ordered state. Experiments were performed at higher frequencies and higher fields as compared to Ref.\onlinecite{Prozorova} as well as at the frequencies from 1 to 20 GHz, thus  closing the gap between NMR and AFMR experiments. Experimental data were described by Andreev-Marchenko theory.\cite{Marchenko} Hysteresis of resonance absorption, which is presumably connected to the domain structure of ordered state in \garnet{}, was detected, domain structure was found to recover slowly once the field is turned off.

\section{Experimental results and discussion}
\subsection{Samples and experimental details}

We have used the same single crystals as in Refs.\onlinecite{Tikhonov}, \onlinecite{TikhonovJL}. For low frequency measurements (below 25 GHz) we have used the $\approx 20$ mg sample, which was previously oriented for NMR experiments with the accuracy of about $1^\circ$. For higher frequency measurements we cut smaller samples (with the mass about 1 mg). These samples were oriented using X-ray diffractometer BRUKER APEX II. Precision of sample orientation transfer from diffractometer to ESR spectrometer was about $3^\circ - 5^\circ$.

To study AFMR in \garnet{} we used set of transmission type spectrometers. Microwave power transmitted through the cavity with the sample was registered as a function of slowly changing magnetic field, microwave frequency remained constant during the field scan. To cover frequency range from 1 to 125 GHz we used several spectrometers differing by the details of cavity geometry and used waveguides.

The main part of results was obtained at $T=1.8$K. Low frequency measurements (1-5 GHz) were performed at the temperatures down to 1.3 K.

\subsection{AFMR frequency-field dependence}
\begin{figure}[htbp]
  \centering
  \includegraphics[width=0.95\columnwidth]{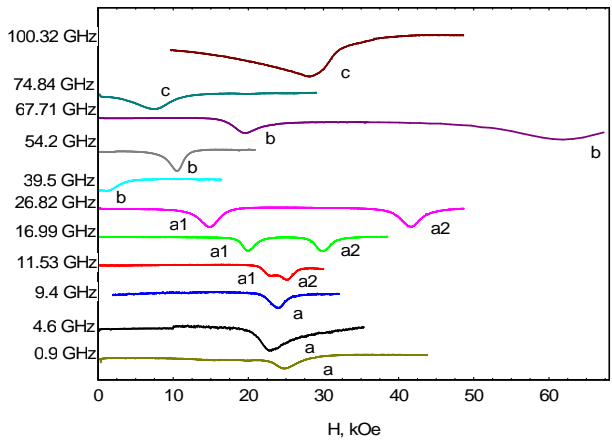}
  \caption{Resonance absorption spectra in \garnet{} for different frequencies in orientation $\vect{H}\parallel[100]$ at the temperature $T=1.8$ K. Indices 'a', 'b', 'c' correspond to different AFMR modes (see Fig.\ref{fig2}).}
\label{fig1}
\end{figure}

\begin{figure}[htbp]
  \centering
  \includegraphics[width=0.9\columnwidth]{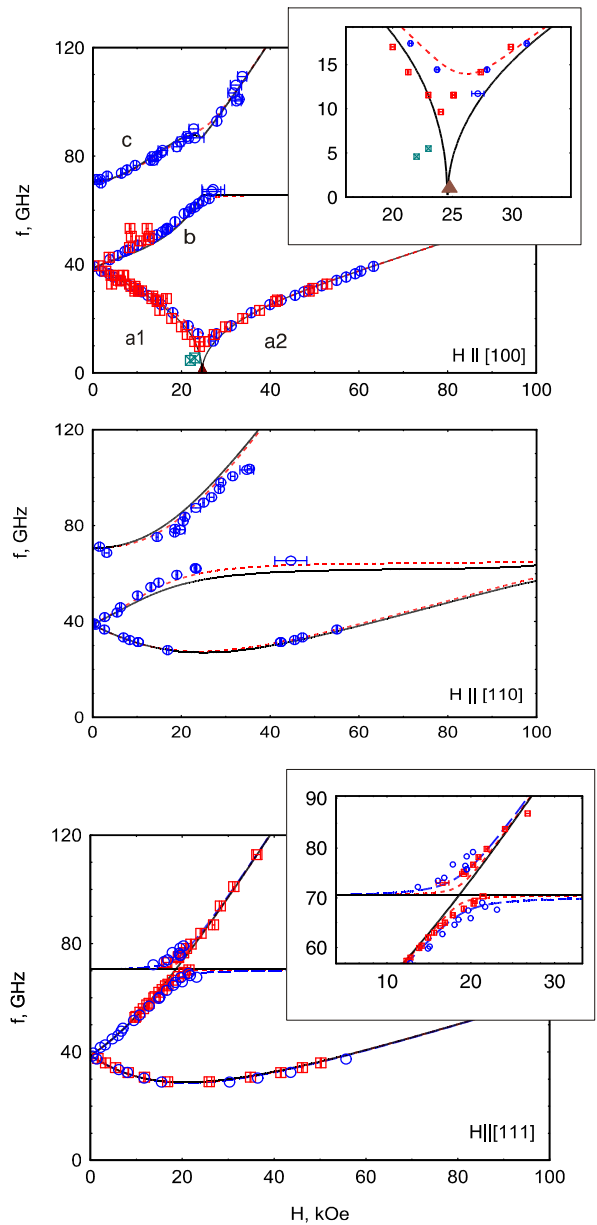}
  \caption{Frequency-field dependence of AFMR for three orientations: $\vect{H}\parallel[100]$, $\vect{H}\parallel[111]$, $\vect{H}\parallel[110]$ at temperature 1.8 K. Solid and dashed lines show results of numeric calculations for exact field orientation and for the field deviated by $5^\circ$ from the exact orientation. For $\vect{H}\parallel[111]$ results of numeric calculations for the field deviated by $10^\circ$ from the exact orientation are also shown. The insets show zoomed fragments of frequency-field dependences of AFMR for $\vect{H}\parallel[100]$ and $\vect{H}\parallel[111]$ orientations.}
\label{fig2}
\end{figure}

The examples of absorption spectra for $\vect{H}\parallel[100]$ are shown in Fig.\ref{fig1}. These data allow to plot frequency-field dependence of AFMR in \garnet{} for given orientation. The frequency-field dependences of AFMR were obtained in this way for three different field orientations $\vect{H}\parallel[100]$, $\vect{H}\parallel[111]$ and $\vect{H}\parallel[110]$ (Fig.\ref{fig2}). There are three modes of AFMR with two gaps about 40 and 70 GHz in all three orientations. At $\vect{H}\parallel[100]$ one of the AFMR modes softens in the field about 24 kOe, marking the spin-reorientation transition. No traces of optical modes were detected in the studied orientations at the frequencies up to 140 GHz and at the fields up to 60 kOe.

Measured frequency-field dependences of AFMR were described by hydrodynamic theory of Andreev-Marchenko.\cite{Marchenko} According to this theory magnetic structure of planar noncollinear antiferromagnet could be parametrized by two mutually orthogonal unitary antiferromagnetic vectors $\vect{l}_1$, $\vect{l}_2$. These vectors lie in the plane of the planar spin structure. Low energy dynamics of such a magnet can be then described in terms of oscillations of  the mutually orthogonal vectors $\vect{l}_1$ , $\vect{l}_2$  and $\vect{n}=\left[\vect{l}_1\times\vect{l}_2\right]$. This treatment can describe only acoustic spin-waves modes, while high energy optical modes are completely ignored. Lagrangian density for our system is:

\begin{eqnarray}
{\cal L}&=&T-U=\frac{I}{2}({\bf\dot{l}}_{1}+\gamma[{\bf{l}}_{1}\times {\bf{H}}])^{2}+\nonumber \\ 
&&+\frac{I}{2}({\bf\dot{l}}_{2}+\gamma[{\bf{l}}_{2}\times {\bf{H}}])^{2}+\frac{I{'}}{2}({\bf\dot{n}}++\gamma[{\bf{n}}\times {\bf{H}}])^{2}-\nonumber \\ 
&&-\lambda\left[\frac{2}{\sqrt{3}}(l_{1x}l_{2x} - l_{1y}l_{2y})+{l_{1z}}^{2} - {l_{2z}}^{2}\right],\label{eqn:lagr}  
\end{eqnarray}

The last term in this equation is an anisotropy energy, its exact form depends on crystal symmetry, we use the same form of anisotropy energy as Ref.\onlinecite{Udalov}. Parameters $I, I{'}$ are dynamic constants, related to static susceptibility as $\chi{'}=2\gamma^{2}I$, $\chi=2\gamma^{2}(I+I{'})$, here $\chi{'}$ is a susceptibility for the field applied parallel to the direction $\vect{n}=[\vect{l}_1\times\vect{l}_2]$ (i.e., perpendicular to the plane of the spin structure) and $\chi$ is a susceptibility for the field applied perpendicular to the $\vect{n}$ direction (i.e., within the plane of spin structure). To find oscillations eigenfrequencies, it is necessary to find equilibrium position for these three vectors ($\vect{l}_1$ , $\vect{l}_2$  and $\vect{n}$) in applied magnetic field and to linearize dynamics equations (Euler-Lagrange equations) in equilibrium state vicinity. In case of \garnet{} it is impossible to solve this problem in analytical way for arbitrary field direction. Thus we used numerical algorithm described in Ref.\onlinecite{Glazkov} to calculate frequency-field dependence of AFMR. Results of modeling are shown in Fig. \ref{fig2}. Model parameters for the curves in Fig.\ref{fig2} are:
$\gamma$=17.59$\frac{10^9 \text{rad}}{\text{sec}\cdot \text{kOe}}$  (which is equal to 2.8 GHz/kOe), $\lambda$=1 ${\text{kOe}}^2$, $I$=1.36$\times$10$^{-5}$ $\frac{{\text{kOe}}^2}{(10^9 \text{rad}{\cdot}\text{sec}^{-1})^2}$, $I^{'}$=9.04$\times$10$^{-6}$ $\frac{{\text{kOe}}^2}{(10^9 \text{rad}{\cdot}\text{sec}^{-1})^2}$.  Gyromagnetic ratio is fixed to the purely spin value, parameter $\lambda$ is arbitrarily fixed to unity (scaling of all terms in the Lagrangian does not change oscillation eigenfrequencies), parameters $I, I{'}$ are fitting parameters. These values of model parameters correspond to the gaps in AFMR spectrum $\nu_{10}$=39.5 GHz , $\nu_{20}$=70 GHz and to the spin reorientation field at $\vect{H}\parallel[100]$ equal to  $H_c=23$ kOe. To take into account possible sample disorientation Fig.\ref{fig2} also includes modeled frequency-field dependences for the field slightly deviated from the exact orientation.

\subsection{Study of interaction between electron and nuclear modes of spin precession}
Theoretical modeling considering only electron precession predicts zeroing of AFMR frequency at the field of spin-reorientation transition. However, hyperfine interaction with nuclear subsystem leads to the repulsion of electron and nuclear modes of spin precession. This repulsion could be estimated theoretically \cite{Udalov}: AFMR frequency at the transition field subject to hyperfine interaction is $\nu_{e} \approx [\gamma_{e}/\gamma_{n}\sqrt{\chi_{n}/\chi_{e}}]\nu_{n}$, where $\gamma_{e}=2.8$ GHz/kOe, $\gamma_n=1.06$ MHz/kOe and $\chi_e=0.3$ emu/mol, $\chi_n=5 \times 10^{-7}$ emu/mol are gyromagnetic ratios and magnetic susceptibilities for electron and nuclear subsystems, and $\nu_n\approx600$ MHz is NMR frequency, these values correspond to T=1.2 K. \cite{TikhonovJL} This estimation yields $\nu_e\approx 2.1$ GHz and expected repulsion of NMR and AFMR modes at 1.2 K is thus about 1.5 GHz.

Examples of absorption spectra and resonance absorption fields for $\vect{H}\parallel[100]$, where spin-reorientation transition is observed, are shown on Fig.\ref{fig1} and Fig.\ref{fig2}. Resonance absorption was observed in the vicinity of the transition field at the frequencies down to 900 MHz. Two components of this signal (corresponding to falling and rising branches of AFMR) were resolvable above 10 GHz. Below 5 GHz absorption signal was very sensitive to sample disorientation: it vanished for the field deviated by $3^\circ$ from the exact orientation.

Zero-field frequency of ${}^{55}$Mn NMR is\cite{TikhonovJL} about 600 MHz, in our AFMR experiment we observed absorption at spin-reorientation field at the frequency of 930 MHz. In other words, apparent repulsion of modes is about 400 MHz, which is considerably less than the estimation made above. This difference between experimental and theoretical values could be explained by big inherent linewidth of antiferromagnetic resonance absorption: half-width of AFMR line near field of softening is about 0.8 kOe (modes 'a1' and 'a2' on Fig.\ref{fig1}), which yields (accounting for frequency-field dependence slope) frequency half-width about 1.5 GHz. Thus, absorption signal at 930 MHz could be, in fact, observed at the wing of resonance curve, shifted due to hyperfine interaction by 1.5 -– 2.0 GHz above the NMR frequency. Our results allow to make only upper estimate for the repulsion of electron and nuclear modes: at 1.8 K it does not exceed 5 GHz (three-fold inherent half-width of AFMR line).

\subsection{Observation of hysteresis of absorption at weak fields}
\begin{figure}[htbp]
  \centering
  \includegraphics[width=0.95\columnwidth]{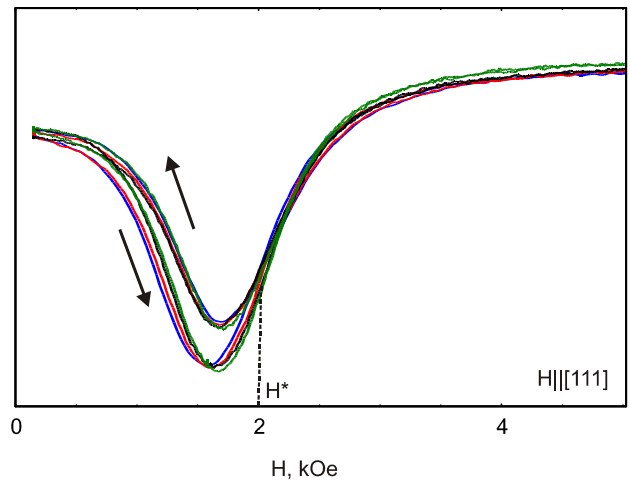}
  \caption{Example of hysteresis of the absorption signal for orientation $\vect{H}\parallel[111]$ at frequency 37.6 GHz and temperature 1.8 K. Data were measured at different field sweep rates: 2 kOe/min (green and black curves), 6 kOe/min (red curve) and 10 kOe/min (blue curve). Dashed line marks monodomenization field $H^*$. Arrows show field sweep directions.}
\label{fig3}
\end{figure}

In the ordered state of \garnet{} spins lay in a plane, which is perpendicular to [111]-like direction. There are four  equivalent orientations of spin plane for the cubic crystal: $\vect{n}\parallel|[111]$, $[-1-11]$, $[1-1-1]$, $[-11-1]$, which could lead to domain structure formation. Since AFMR resonance field depends on orientation of magnetic field with respect to the ordered structure, presence of several domains should lead to the appearance of the additional absorption lines. However, we have not detected additional absorption while we were measuring AFMR spectra. Meanwhile at frequencies near gaps hysteresis of absorption signal was detected (fig. \ref{fig3}). We have observed it for all studied field orientations ($\vect{H}\parallel[100]$, $\vect{H}\parallel[111]$, $\vect{H}\parallel[110]$). We have checked that the observed hysteresis did not depend on field sweep speed, microwave power level and on the details of sample mounting.
 
Observed absorption signal depends on sample history. Absorption spectrum recorded at rising field right after zero-field cooling differs both from the spectrum recorded at decreasing field and from the consequentive absorption spectra recorded at rising field. Absorption spectra recorded at decreasing field are reproduced independently of sample history (including field cooling).

For all sample orientations absorption spectra recorded at rising and decreasing fields coincides above certain threshold field $H^*$. The value of this field depends on orientation: at 1.8K for $\vect{H}\parallel[100]$ $H^*=(7.5\pm0.5)$ kOe, for $\vect{H}\parallel[110]$ $H^*=(3.5\pm0.5)$ kOe, for $\vect{H}\parallel[111]$ $H^*=(2.0\pm0.2)$ kOe  at T=1.8 K. We have not studied threshold field temperature dependence systematically, but for $\vect{H}\parallel[100]$ at $T=4.2$K threshold field slightly drops to $H^*=(6.0\pm0.2)$ kOe. If the field sweep direction is changed before the threshold field is reached the hysteresis persists but 'the  hysteresis loop' is reduced.

In some experiments (near gap frequencies) we observed that upon completion of the field sweep to the field above the threshold field and back to zero field microwave signal transmitted through the cavity with the sample differs from the initial microwave power level and slowly recovers to the initial power level afterwards. The typical recovery time in zero field was about 20 seconds.

The observed hysteresis could be connected with transformation of domain structure in the magnetic field. According to the results of our experiments, monodomenization takes place at the field about 10 kOe. This fact differs from NMR nonlinear radiofrequency absorption reported in Ref.\onlinecite{Tikhonov}, which was explained as oscillations of the domain boundaries --- in NMR experiment the effect of nonlinear absorption was observed up to reorientation transition field (24 kOe).

\section{Conclusions}

We have measured frequency-field dependences of antiferromagnetic resonance for \garnet{} at frequencies 1-120 GHz and fields up to 6 T. These dependences are well described theoretically by hydrodynamic theory of Andreev-Marchenko. Our data complement previously measured AFMR frequency-field dependences \cite{Prozorova} and contain new results. We have significantly broadened range of measurements as compared to previous studies and we have studied in details AFMR near spin-reorientation transition field, where repulsion of electron and nuclear modes of spin precession takes place. Signal of absorption observed around 1 GHz is, probably, due to big inherent width of the AFMR line. We have observed hysteresis of resonance absorption, it is connected with transformation of sample domain structure. Meanwhile sample monodomenization was observed above certain threshold field, the value of threshold field is orientation dependent.

\acknowledgements

The authors are grateful to Prof. B. V. Mill (MSU) for providing the samples, Prof. A. M. Tikhonov for discussion in the process. Also, the authors thank Prof. A. I. Smirnov and Prof. L. E. Svistov for valuable comments and support in the work.

Work was supported by RFBR grant No 16-02-00688, RSCF grant No 17-12-01505 and also by RAS Program of support of fundamental studies 'Electron spin resonance, spin-dependent effects and spin technologies'. Work in HSE (Y.V.K. and V.N.G.) was supported by Program of fundamental studies of HSE.


\begin{references}
\bibitem{Marchenko}  A.F. Andreev, V.I. Marchenko, Sov. Phys. Usp. \textbf{130}, 39
(1980)
\bibitem{Prandl} W. Prandl,  phys. stat. sol. (b) \textbf{55}, K159 (1973)
\bibitem{Plahtii} A. Gukasov, V. P. Plakhty, B. Dorner, S. Yu. Kokovin, V. N. Syromyatnikov, O. P. Smirnov and Yu. P. Chernenkov, J.Phys.: Condens. Matter \textbf{11}, 2869 (1999)
\bibitem{Kazey} Z. A. Kazei, N. P. Kolmakova, M. V. Levanidov, B. V. Mill, and V. I. Sokolov, JETP \textbf{65}, 1283 (1987) [Zh. Eksp. Teor. Fiz. \textbf{92}, 2277 (1987)]
\bibitem{Tikhonov}  A. M. Tikhonov, N. G. Pavlov, JETP Letters \textbf{99}, 229 (2014) [Pis'ma v ZETF 99, 255 (2014)]
\bibitem{Prozorova} L. A. Prozorova, V. I. Marchenko, Yu. V. Krasnyak, JETP Letters \textbf{41}, 637 (1985) [Pis'ma Zh.Eksp.Teor.Fiz \textbf{41},522 (1985)]
\bibitem{TikhonovJL}  A. M. Tikhonov, N. G. Pavlov, O. G. Udalov   JETP Letters \textbf{96}, 517 (2012) [Pis'ma Zh.Eksp.Teor.Fiz. \textbf{96}, 568  (2014)]
\bibitem{Udalov} O. G. Udalov JEPT \textbf{113}, 490 (2011) [Zh.Eksp.Teor.Fiz \textbf{140}, 561 (2011)]
\bibitem{Glazkov}  V. Glazkov, T. Soldatov, Yu. Krasnikova, Appl. Magn. Reson. \textbf{47}, 1069 (2016)


\end{references}
\end{document}